\documentclass
[superscriptaddress,secnumarabic,amssymb,amsmath,nobibnotes,aps,prd,showkeys,showpacs,nofootinbib,onecolumn]{revtex4}%
\usepackage{graphicx}
\usepackage{epsf}
\usepackage{bm}
\usepackage{amsmath}
\usepackage{amsfonts}
\usepackage{amssymb}%
\providecommand{\U}[1]{\protect\rule{.1in}{.1in}}

\newcommand{\be}{\begin{equation}}
\newcommand{\ee}{\end{equation}}

\newcommand{\mincir}{\raise
-3.truept\hbox{\rlap{\hbox{$\sim$}}\raise4.truept\hbox{$<$}\ }}
\newcommand{\magcir}{\raise
-3.truept\hbox{\rlap{\hbox{$\sim$}}\raise4.truept\hbox{$>$}\ }}

\ifx\pdfoutput\relax\let\pdfoutput=\undefined\fi
\newcount\msipdfoutput
\ifx\pdfoutput\undefined\else
\ifcase\pdfoutput\else
\ifx\paperwidth\undefined\else
\ifdim\paperheight=0pt\relax\else\pdfpageheight\paperheight\fi
\ifdim\paperwidth=0pt\relax\else\pdfpagewidth\paperwidth\fi
\fi\fi\fi
\begin{document}
\title{Szekeres Universes with Homogeneous Scalar Fields}
\author{John D. Barrow}
\email{J.D.Barrow@damtp.cam.ac.uk}
\affiliation{DAMTP, Centre for Mathematical Sciences, University of Cambridge, Wilberforce
Rd., Cambridge CB3 0WA, UK}
\author{Andronikos Paliathanasis}
\email{anpaliat@phys.uoa.gr}
\affiliation{Instituto de Ciencias F\'{\i}sicas y Matem\'{a}ticas, Universidad Austral de
Chile, Valdivia, Chile}
\affiliation{Department of Mathematics and Natural Sciences, Core Curriculum Program,
Prince Mohammad Bin Fahd University, Al Khobar 31952, KSA}
\affiliation{Institute of Systems Science, Durban University of Technology, PO Box 1334,
Durban 4000, RSA}

\begin{abstract}
We consider the existence of an \textquotedblleft inflaton\textquotedblright%
\ described by an homogeneous scalar field in the Szekeres cosmological
metric. The gravitational field equations are reduced to two families of
solutions which describe the homogeneous Kantowski-Sachs spacetime and an
inhomogeneous FLRW(-like) spacetime with spatial curvature a constant. The
main differences with the original Szekeres spacetimes containing only
pressure-free matter are discussed. We investigate the stability of the two
families of solution by studying the critical points of the field equations.
We find that there exist stable solutions which describe accelerating
spatially-flat\ FLRW geometries.

\end{abstract}
\maketitle
\date{\today}

\section{Introduction}

The main mechanism to explain the isotropization of the observable part of the
universe today from a general set of initial conditions by means of an early
period of accelerated expansion, the so-called inflationary epoch, is often
based on the existence of an explicit or effective scalar field dubbed the
\textquotedblleft inflaton\textquotedblright\ \cite{guth}. The scalar field
temporarily dominates the expansion dynamics and drives them towards a locally
isotropic and homogeneous form that leaves only very small residual
anisotropies at the end of a brief inflaton-dominated period. Quantum
fluctuations are also processed by the period of inflation and can manifest
themselves as density and gravitational-wave inhomogeneities at late times.
Consequently, pre-inflationary anisotropies could have played an important
role in the evolution of the universe.

The Bianchi class of spatially homogeneous cosmologies contains several
important cosmological models that have been used for the discussion of
anisotropies of primordial universe and for its evolution towards the observed
isotropy of the present epoch \cite{Mis69,jacobs2,collins,JB1,JB2}. \ Detailed
analysis of the Einstein field equations for Bianchi cosmologies with a
cosmological constant \cite{wald}, and with a changing scalar field have shown
that isotropic Friedmann--Lema\^{\i}tre--Robertson--Walker (FLRW) attractor
solutions exist for specific initial conditions when the scalar field
potential has a large positive value \cite{heu}. In the case where the scalar
field potential is exponential, exact solutions can be found and algebraic
conditions that guarantee isotropization of ever-expanding homogeneous Bianchi
universes have been derived in \cite{coley1,coley2} by studying the critical
points of the field equations. Similar results for the Kantowski-Sachs
spacetime and Bianchi types I and V had been found earlier by Burd et al. in
\cite{bb1} although these spacetimes are not subject to the usual no-hair
theorems because they have positive 3-curvature.

On the other hand, it has been found that the existence of small
inhomogeneities in the spacetime does affect necessary the existence of
inflation while as it has been shown by Turner et al. \cite{f3} that some
homogeneous models which start from inhomogeneous models can become
anisotropic in the future. However, because of the inflation that will happen
in an exponentially distant time in the future, and in the present era the
models to be still homogeneous up to very small ($O(10^{-5})$) metric perturbations.

An important family of inhomogeneous analytic spacetimes are the Szekeres
spacetimes \cite{szek0}. They belong to the class of 'Silent' universes where
information does not propagate via gravitational or sound waves. This requires
the magnetic part of the Weyl tensor to be zero and the total matter source to
be described by an {irrotational isotropic dust fluid. An important property
of the Szekeres spacetimes is that they do not admit isometries, hence these
spacetimes }have been characterized as \textquotedblleft
partially\textquotedblright\ locally rotational spacetimes {\cite{musta}}.
Furthermore, it was found that the Szekeres system remains invariant with
respect quantum corrections \cite{ansz}. Although they possess no Killing
symmetries, Szekeres universes are special in other ways because the matter
distribution has a dipolar character \cite{shawB}. The missing changing
quadrupole ensures that there is no gravitational radiation emission from the
inhomogeneously moving dust \cite{bon}. They have Newtonian analogues
\cite{Gol} and are the general relativistic generalisation
\cite{grpan1,grpan2,grpan3} of the newtonian 'pancake' approximation
introduced by Zeldovich \ \cite{zeldpan}.

Szekeres spacetimes are important because they have applications in many areas
of gravitational physics and cosmology \cite{sz1,sz2,sz3,sz4,sz5}.
Inhomogeneous Szekeres exact solutions with a cosmological constant were
derived by Barrow et al. in \cite{barcc}, and others with a general
time-dependent pressure are given in \cite{szafron,Bgron}. The solutions of
\cite{barcc} have been found to be inhomogeneous generalizations of the de
Sitter spacetime and were the first analytic solutions of inhomogeneous
expanding inflationary spacetimes\footnote{Szekeres metrics in 2+1 dimensional
spacetimes were found in ref \cite{BST}.}. These provide a basis for the study
of non-linear inhomogeneities in inflation.

Here, we consider the Szekeres metric with a self-interacting homogeneous
scalar field. The scalar field is able to describe an inflaton\ field and the
FLRW limit exist for the resulting solutions of the field equations. For the
conditions required for a FLRW limit see \cite{krasinski}. We know that in
general the Szekeres diagonal form of the metric requires any diagonal
pressure in the energy-momentum tensor to depend on time but not on space. In
particular this is why exact solutions are found with dust and with dust and a
cosmological constant. In the case of the homogeneous scalar field the
pressure is restricted to being a function only of the time while the metric
may depend on the time and space coordinates.

As in the case of the Szekeres system, with or without the cosmological
constant, we find two sets of solutions which correspond to the (a)
Kantowski-Sachs family and to the (b) FLRW family, of spacetimes. However, the
Kantowski-Sachs family of solutions in the presence of the homogeneous scalar
field turns out to be spatially homogeneous. This is not true for the second
family of solutions. Specifically, the second family are inhomogeneous
FLRW-like spacetimes in which the \textquotedblleft spatial
curvature\textquotedblright\ does not depend upon any variable, just as in the
FLRW models. The plan of the paper is as follows.

In Section \ref{section2} we define our model which is a Szekeres metric with
a homogeneous scalar field. The two different families of solutions of our
dynamical system are presented in Section \ref{section3}. The dynamical
analysis of the critical points of the Szekeres system with the scalar field
is performed in Section \ref{section4a}. Finally, in Section \ref{section5} we
draw some conclusions.

\section{Szekeres system with a homogeneous scalar field}

\label{section2}

\qquad In the context of general relativity we consider the following
four-dimensional spacetime first considered by Szekeres \cite{szek0}:%
\begin{equation}
ds^{2}=-dt^{2}+e^{2A}dr^{2}+e^{2B}\left(  dy^{2}+dz^{2}\right)  ,
\label{sf.00}%
\end{equation}
where $A=A\left(  t,r,y,z\right)  $ and $B=B\left(  t,r,y,z\right)  $ are to
be determined by the Einstein field equations.

The energy-momentum tensor, $T_{\mu\nu}$, is assumed to be given by the
expression
\begin{equation}
T_{\mu\nu}=T_{\mu\nu}^{\left(  D\right)  }+T_{\mu\nu}^{\left(  \phi\right)  },
\label{sf.02}%
\end{equation}
where $T_{\mu\nu}^{\left(  D\right)  }=\rho u_{\mu}u_{\nu}$ describes a
pressureless fluid source (dust) in which $u^{\mu}=\delta_{t}^{\mu}$ is the
comoving 4-velocity.

We take $T_{\mu\nu}^{\left(  \phi\right)  }$ to be the energy-momentum tensor
of a scalar field with potential,$~V\left(  \phi\right)  ,$ defined as usual
by
\begin{equation}
T_{\mu\nu}^{\left(  \phi\right)  }=\frac{1}{2}\left[  \phi_{,\mu}\phi_{,\nu
}-\frac{1}{2}g_{\mu\nu}\left(  \phi^{,\mu}\phi_{,\mu}+2V(\phi)\right)
\right]  . \label{sf.03}%
\end{equation}

The gravitational field equations are%
\begin{equation}
G_{\mu\nu}=T_{\mu\nu}^{\left(  D\right)  }+T_{\mu\nu}^{\left(  \phi\right)  },
\label{sf.04}%
\end{equation}
plus the separate conservation equations%
\begin{equation}
T_{~\ ~~~\ ~~~~;\nu}^{\left(  D\right)  \mu\nu}=0~~,~T_{~\ ~~~\ ~~~~;\nu
}^{\left(  \phi\right)  \mu\nu}=0. \label{sf.05}%
\end{equation}

The latter dynamical system without the scalar field describes the original
Szekeres system \cite{szek0}. \ By assuming that the solution of the field
equations has a FLRW limit it follows that $p_{\phi}=p_{\phi}\left(  t\right)
~$\cite{krasinski}, in which $p_{\phi}=\frac{1}{3}T_{\mu\nu}^{\left(
\phi\right)  }\left(  g^{\mu\nu}+u^{\mu}u^{\nu}\right)  .$

Hence, in order for the latter to be true in the limit of the scalar field
becoming a stiff perfect fluid, that is, $\phi=\phi\left(  t\right)  ,~$we
consider that $\phi=\phi\left(  t\right)  $. Then the continuity equation
$T_{~\ ~~~\ ~~~~;\nu}^{\left(  \phi\right)  \mu\nu}=0$ provides the
differential equation%
\begin{equation}
\frac{d^{2}\phi}{dt^{2}}+\left(  \left(  \frac{\partial A}{\partial t}\right)
+2\left(  \frac{\partial B}{\partial t}\right)  \right)  \left(  \frac{d\phi
}{dt}\right)  +\frac{dV}{d\phi}=0, \label{sf.06}%
\end{equation}
from which it follows that
\begin{equation}
\exp\left(  A\left(  t,r,y,z\right)  \right)  =a\exp\left(  F\left(
r,y,z\right)  -2B\left(  t,r,y,z\right)  \right)  . \label{sf.07}%
\end{equation}
Thus, the spacetime metric (\ref{sf.00}) is simplified to%
\begin{equation}
ds^{2}=-dt^{2}+a^{2}\left(  t\right)  e^{2F\left(  r,y,z\right)  -4B\left(
t,r,y,z\right)  }dr^{2}+e^{2B\left(  r,y,z\right)  }\left(  dy^{2}%
+dz^{2}\right)  . \label{sf.08}%
\end{equation}

We continue with the reduction of the field equations to a set of ordinary
differential equations with respect to the comoving proper time parameter $t$,
and define explicitly the geometry of the spacetime.

\section{Families of spacetimes}

\label{section3}

In a similar way to the case without the scalar field, the solution of the
field equations is given by the two particular families of solutions where (A)
$\frac{\partial B}{\partial r}=0$ and (B) $\frac{\partial B}{\partial r}\neq0$.

\subsection{Kantowski-Sachs family: $\frac{\partial B}{\partial r}=0$}

In the case in which $\frac{\partial B}{\partial r}=0$, the line element
reduces to the Kantowski-Sachs spacetime,%
\begin{equation}
ds^{2}=-dt^{2}+a^{2}\left(  t\right)  dr^{2}+\beta^{2}\left(  t\right)
e^{2C\left(  y,z\right)  }\left(  dy^{2}+dz^{2}\right)  , \label{sf.11a}%
\end{equation}
where
\begin{equation}
C\left(  y,z\right)  =-2\ln\left(  c_{1}uv+c_{2}u+c_{3}v+c_{4}\right)  \,
\label{sf.12}%
\end{equation}
and the new complex variables $\left\{  u,v\right\}  $ are defined as
\begin{equation}
y=u+v,~z=i\left(  u-v\right)  , \label{sf.13}%
\end{equation}
while the constants $c_{1}\rightarrow c_{4}$ are related to the curvature,
$K,~$of the two-dimensional surface $\left\{  y-z\right\}  $ as follows
\begin{equation}
c_{1}c_{4}-c_{2}c_{3}=K. \label{sf.14}%
\end{equation}

Consequently, the field equations (\ref{sf.09})-(\ref{sf.11}) are those for
the Kantowski-Sachs spacetime with a scalar field. That is, the field
equations (\ref{sf.04})-(\ref{sf.05}) are reduced to the following system of
ordinary differential equations
\begin{equation}
\frac{2}{a\beta}\left(  \frac{da}{dt}\right)  \left(  \frac{d\beta}%
{dt}\right)  +\frac{1}{\beta^{2}}\left(  \frac{d\beta}{dt}\right)
^{2}+\left(  \frac{d\phi}{dt}\right)  ^{2}+2V\left(  \phi\right)  +\rho
_{0}a^{-1}\beta^{-2}+\frac{K}{\beta^{2}}=0 \label{sf.09}%
\end{equation}%
\begin{equation}
\frac{1}{a}\left(  \frac{d^{2}a}{dt^{2}}\right)  +\frac{1}{\beta}\left(
\frac{d^{2}\beta}{dt^{2}}\right)  +\frac{1}{a\beta}\left(  \frac{da}%
{dt}\right)  \left(  \frac{d\beta}{dt}\right)  +\frac{1}{2}\left(  \frac
{d\phi}{dt}\right)  ^{2}-2V\left(  \phi\right)  =0 \label{sf.10}%
\end{equation}%
\begin{equation}
-2\left(  \frac{d^{2}\beta}{dt^{2}}\right)  -\frac{1}{\beta^{2}}\left(
\frac{d\beta}{dt}\right)  ^{2}+\frac{K}{\beta^{2}}-\frac{1}{2}\left(
\frac{d\phi}{dt}\right)  ^{2}+V\left(  \phi\right)  =0 \label{sf.11}%
\end{equation}
plus the conservation equation (\ref{sf.06}).

At this point it is interesting that the inhomogeneous spacetime (\ref{sf.08})
reduces to the homogeneous Kantowski-Sachs element and not to the
inhomogeneous Kantowski-Sachs(-like) as in the case without the scalar field
\cite{szek0,barcc,szafron}. Hence, we can infer that the existence of the
homogeneous scalar field provides an homogeneous anisotropic universe. The
latter property is not true for the second family of solutions.

There are a few analytic solutions for the field equations (\ref{sf.09}%
)-(\ref{sf.11}). For instance, a solution without the dust fluid term and with
zero potential is presented in \cite{xanth}; while \cite{dabrowski} gives
exact solutions are presented for string cosmologies. Recall that when $K=0$,
Kantowski-Sachs spacetime reduces to the Bianchi I. For the Bianchi I
spacetime analytic solutions with a scalar field without a matter source are
given in \cite{bia1,bia2,bia3}. Last, the generic vacuum solution for the
Kantowski-Sachs spacetime can be found in \cite{chr2}

\subsection{FLRW family: $\frac{\partial B}{\partial r}\neq0$}

For the second family of solutions, the Szekeres line element reduces to
%
\begin{equation}
ds^{2}=-dt^{2}+a^{2}\left(  t\right)  \left(  \left(  \frac{\partial C\left(
r,y,z\right)  }{\partial r}\right)  ^{2}dr^{2}+e^{2C\left(  r,y,z\right)
}\left(  dy^{2}+dz^{2}\right)  \right)  , \label{sf.15}%
\end{equation}
where the spatial function $C\left(  r,y,z\right)  ~$is given by
\begin{equation}
C\left(  r,y,z\right)  =-\ln\left(  \gamma_{1}\left(  r\right)  uv+\gamma
_{2}\left(  r\right)  u+\gamma_{3}\left(  r\right)  v+\gamma_{4}\left(
r\right)  \right)  \,. \label{sf.16}%
\end{equation}
The unctions $\gamma_{1}\left(  r\right)  \rightarrow\gamma_{4}\left(
r\right)  $ are related by%
\begin{equation}
\gamma_{1}\left(  r\right)  \gamma_{4}\left(  r\right)  -\gamma_{2}\left(
r\right)  \gamma_{4}\left(  r\right)  =k, \label{sf.17}%
\end{equation}
in which $k$ is a constant and not a function of $r$, as was the case without
the scalar field.

Furthermore, the scale factor $a\left(  t\right)  $ and the scalar field
$\phi\left(  t\right)  $ satisfy Friedmann's equations
\begin{equation}
-\frac{2}{a}\frac{d^{2}a}{dt^{2}}-\left(  \frac{da}{dt}\right)  ^{2}%
+ka^{-2}-\frac{1}{2}\left(  \frac{d\phi}{dt}\right)  ^{2}+2V\left(
\phi\right)  =0, \label{sf.18}%
\end{equation}%
\begin{equation}
-\frac{3}{a^{2}}\left(  \frac{da}{dt}\right)  ^{2}+\frac{3}{2}ka^{-2}+\left(
\frac{d\phi}{dt}\right)  ^{2}+2V\left(  \phi\right)  +\rho=0 \label{sf.19}%
\end{equation}
and the conservation equation (\ref{sf.06}).

The main difference to the case without the scalar field is that the spatial
curvature $k$ is a constant and not a function of $r$. However, the spacetime
metric (\ref{sf.15}) remains inhomogeneous as in \cite{szek0,barcc,szafron}.

There are various analytical solutions for the field equations (\ref{sf.18}),
(\ref{sf.19}) with or without the dust fluid, and with zero or nonzero spatial
curvature, for instance see \cite{sf1,sf1a,sf1b,sf1c,sf2,sf3,sf4,sf5} while
some analytical solutions with application in inflation are presented in
\cite{sf6,sf7} and references therein.

In order to understand the evolution of the Szekeres spacetime with the scalar
field and study the stability of the family of solutions that we have
presented, in the next Section we perform an analysis of the critical points
of the Einstein field equations.

\section{Dynamical evolution}

We now study the dynamical evolution of the system using the covariant
kinematic variables of Ehlers and Ellis \cite{silent1,ellis1,ellis2}. The
Einstein equations for silent universes with pressure $p,$ are equivalent to
the following system for the density, pressure, volume expansion rate
$\theta,~$shear scalar $\sigma,$ and scalar electric part of the Weyl tensor,
$\mathcal{E}$:%

\begin{align}
\frac{d\rho}{dt}+\theta\left(  \rho+p\right)   &  =0,~\label{ss.01}\\
\frac{d\theta}{dt}+\frac{\theta^{2}}{3}+6\sigma^{2}+\frac{1}{2}\left(
\rho+3p\right)   &  =0,\label{ss.02}\\
\frac{d\sigma}{dt}-\sigma^{2}+\frac{2}{3}\theta\sigma+\mathcal{E}  &
=0,\label{ss.04}\\
\frac{d\mathcal{E}}{dt}+3\mathcal{E}\sigma+\theta\mathcal{E}+\frac{1}%
{2}\left(  \rho+p\right)  \sigma &  =0, \label{ss.05}%
\end{align}
with the constraint
\begin{equation}
\frac{\theta^{2}}{3}-3\sigma^{2}+\frac{^{\left(  3\right)  }R}{2}=\rho,
\label{ss.06}%
\end{equation}
where $^{\left(  3\right)  }R$ denotes the curvature of the three-dimensional hypersurfaces.

The total fluid still comprises a pressureless perfect fluid (dust) and a
minimally homogeneous scalar field with self-interaction potential $V(\phi)$:%

\begin{equation}
\rho=\rho_{D}+\rho_{\phi}=\rho_{D}+\left(  \frac{1}{2}\dot{\phi}^{2}+V\left(
\phi\right)  \right)  , \label{ss.07}%
\end{equation}%
\begin{equation}
p=p_{\phi}=\frac{1}{2}\dot{\phi}^{2}-V\left(  \phi\right)  . \label{ss.08}%
\end{equation}
Since the two fluids are not interacting it follows from equation
(\ref{ss.01}) that
\begin{equation}
\frac{d\rho_{D}}{dt}+\theta\rho_{D}=0, \label{ss.09}%
\end{equation}%
\begin{equation}
\frac{d^{2}\phi}{dt^{2}}+\theta\frac{d\phi}{dt}+V\left(  \phi\right)  _{,\phi
}=0. \label{ss.10}%
\end{equation}

The field equations (\ref{ss.02})-(\ref{ss.06}) become

\bigskip%

\begin{align}
\frac{d\theta}{dt}+\frac{\theta^{2}}{3}+6\sigma^{2}+\frac{1}{2}\rho
_{D}+\left(  \left(  \frac{d\phi}{dt}\right)  ^{2}-V\left(  \phi\right)
\right)   &  =0,\label{ss.11}\\
\frac{d\sigma}{dt}-\sigma^{2}+\frac{2}{3}\theta\sigma+\mathcal{E}  &
=0,\label{ss.12}\\
\frac{d\mathcal{E}}{dt}+3\mathcal{E}\sigma+\theta\mathcal{E}+\frac{1}%
{2}\left(  \rho_{D}+\left(  \frac{d\phi}{dt}\right)  ^{2}\right)  \sigma &  =0
\label{ss.13}%
\end{align}
and%
\begin{equation}
\frac{\theta^{2}}{3}-3\sigma^{2}+\frac{^{\left(  3\right)  }R}{2}=\rho
_{D}+\frac{1}{2}\left(  \frac{d\phi}{dt}\right)  ^{2}+V\left(  \phi\right)  .
\label{ss.14}%
\end{equation}

In order to proceed with the study of the critical points we define the new
dimensionless variables scaled by appropriate \ powers of the volume Hubble
expansion rate, $\theta:$
\begin{equation}
\Omega_{D}=3\frac{\rho_{D}}{\theta^{2}}~,~\Sigma=\frac{\sigma}{\theta
}~,~\varepsilon=\frac{\mathcal{E}}{\theta^{2}}~,~y\left(  t\right)
=\frac{\sqrt{6}}{2\theta}\left(  \frac{d\phi}{dt}\right)  ~,\text{~}%
z=\frac{3V\left(  \phi\right)  }{\theta^{2}}~\,\text{and }\lambda
=-\frac{V\left(  \phi\right)  _{,\phi}}{V}. \label{ss.15}%
\end{equation}
Moreover, we consider the new independent variable to be $N\left(  t\right)
$, such that $dN\left(  t\right)  =\theta\left(  t\right)  dt$, so now the
field equations (\ref{ss.09})-(\ref{ss.13}) can be rewritten as
\begin{equation}
\frac{d\Omega_{D}}{dN}=\frac{1}{3}\Omega_{D}\left(  \Omega_{D}-1+36\Sigma
^{2}+2\left(  y^{2}-z\right)  \right)  , \label{ss.16}%
\end{equation}%
\begin{equation}
\frac{d\Sigma}{dN}=\frac{\Sigma}{6}\left(  4y^{2}-2z-2+6\Sigma\left(
1+6\Sigma\right)  +\Omega_{D}\right)  -\varepsilon, \label{ss.17}%
\end{equation}%
\begin{equation}
\frac{d\varepsilon}{dN}=\frac{\varepsilon}{3}\left(  4y^{2}-2z-1+9\Sigma
\left(  4\Sigma-1\right)  +\Omega_{D}\right)  -\frac{\Sigma}{6}\left(
2y^{2}+\Omega_{D}\right)  , \label{ss.18}%
\end{equation}%
\begin{equation}
\frac{dy}{dN}=\frac{1}{6}\left(  4y^{3}+\sqrt{6}\lambda z+y\left(
2z-4+36\Sigma^{2}+\Omega_{D}\right)  \right)  , \label{ss.19}%
\end{equation}%
\begin{equation}
\frac{dz}{dN}=\frac{z}{3}\left(  2\left(  2y^{2}-z+1\right)  +36\Sigma
^{2}+\sqrt{6}\lambda y+\Omega_{D}\right)  , \label{ss.20}%
\end{equation}
and%
\begin{equation}
\frac{d\lambda}{dN}=-\sqrt{6}\lambda^{2}y\left(  \Gamma\left(  \lambda\right)
-1\right)  , \label{ss.21}%
\end{equation}
where $\Gamma\left(  \lambda\right)  =\frac{V_{,\phi\phi}V}{V_{,\phi}^{2}}%
$~\cite{cop1}.

We assume that the scalar field potential is purely exponential, $V\left(
\phi\right)  =V_{0}e^{-\sigma\phi}$, so that $\lambda=\sigma$ , and the
resulting dynamical system is reduced from a six-dimensional to a
five-dimension system comprising the differential equations (\ref{ss.16}%
)-(\ref{ss.20}).

The exponential potential captures a very wide range of slow roll potentials,
including power-law inflation and no infltation (for steep exponential) and de
Sitter inflation when the exponent is zero. It does not possess a minimum,
where non-inflationary oscillatory behaviour will occur but no exact solution
will be possible. The exponential potential allows exact solutions in the
homogeneous and isotropic case and so is a strong candidate for exact
solutions in this inhomogeneous situation. It is also conformally related to
important higher-order gravity theories with quadratic lagrangians.

The algebraic equation (\ref{ss.14}), becomes
\begin{equation}
\Omega_{R}=1-y^{2}-z-9\Sigma^{2}-\Omega_{D},\label{ss.22}%
\end{equation}
with $\Omega_{R}=-\frac{3}{2}\frac{^{\left(  3\right)  }R}{\theta^{2}}$.
Moreover, paramters $z$ and $\Omega_{D}$ are positive parameters. The
algebraic equation (\ref{ss.22}) is the one which defines the invariant sets
on specific surfaces. For more details on the invariant sets of Bianchi
cosmologies with a scalar field we refer the reader in \cite{colbook}%
\textbf{.}

Furthermore, at the critical points for the Raychaudhuri equation
(\ref{ss.11}) it follows that%
\begin{equation}
\dot{\theta}=-\frac{1}{6}\theta^{2}\left(  2+4y_{p}^{2}-2z_{p}+36\Sigma
_{p}^{2}+\Omega_{p}\right)  =-\frac{1}{\theta_{0}}\theta^{2}, \label{ss.23}%
\end{equation}
so that the solution for the expansion rate $\theta\left(  t\right)  $ is
\begin{equation}
\theta\left(  t\right)  =\frac{\theta_{0}}{t-t_{0}}. \label{ss.24}%
\end{equation}

\label{section4a}The dynamical system, (\ref{ss.16})-(\ref{ss.20}), admits
sixteen critical points, which form three different families. The first family
(A) admits seven critical points and correspond to solutions of the system
without the scalar field, that is, with $y\left(  A\right)  =z\left(
A\right)  =0\,;$ however, one of the critical points corresponds to the case
with $\Omega_{D}<0$, which means that it is unphysical. In the second family
(B) there are two critical points. At these two points only the kinetic term
of the scalar field contributes in the solution, that is, $z\left(  B\right)
=0=V(\phi),$ so they correspond to solutions with stiff $p=\rho$ perfect
fluid. The remaining seven points correspond to the third family (C) of
solutions in which $y\left(  C\right)  z\left(  C\right)  \neq0.$ However,
given the condition $\Omega_{D}\geq0$ on the density, only five points are
physically acceptable.%

\begin{table}[tbp] \centering
\caption{Critical points of family (A)}%
\begin{tabular}
[c]{ccccccc}\hline\hline
\textbf{Point} & $\left(  \mathbf{\Omega,\Sigma,\varepsilon}\right)  $ &
$\left(  \mathbf{y,z}\right)  $ & \textbf{Physical} & $^{\left(  3\right)
}\mathbf{R}$ & \textbf{Spacetime} & \textbf{Stability}\\\hline
$A_{1}$ & $\left(  1,0,0\right)  $ & $\left(  0,0\right)  $ & Yes & $=0$ &
FLRW (Spatially flat) & Unstable\\
$A_{2}$ & $\left(  0,0,0\right)  $ & $\left(  0,0\right)  $ & Yes & $<0$ &
FLRW (Milne universe) & Unstable\\
$A_{3}$ & $\left(  0,-\frac{1}{3},0\right)  $ & $\left(  0,0\right)  $ & Yes &
\thinspace$=0$ & Bianchi I (Kasner universe) & Unstable\\
$A_{4}$ & $\left(  0,\frac{1}{3},\frac{2}{9}\right)  $ & $\left(  0,0\right)
$ & Yes & \thinspace$=0$ & Bianchi I (Kasner universe) & Unstable\\
$A_{5}$ & $\left(  0,\frac{1}{6},0\right)  $ & $\left(  0,0\right)  $ & Yes &
$<0$ & Kantowski-Sachs & Unstable\\
$A_{6}$ & $\left(  0,-\frac{1}{12},\frac{1}{32}\right)  $ & $\left(
0,0\right)  $ & Yes & $<0$ & Kantowski-Sachs & Unstable\\
$A_{7}$ & $\left(  -3,-\frac{1}{3},\frac{1}{6}\right)  $ & $\left(
0,0\right)  $ & No &  &  & \\\hline\hline
\end{tabular}
\label{family1}%
\end{table}%
%

\begin{table}[tbp] \centering
\caption{Critical points of family (B)}%
\begin{tabular}
[c]{ccccccc}\hline\hline
\textbf{Point} & $\left(  \mathbf{\Omega,\Sigma,\varepsilon}\right)  $ &
$\left(  \mathbf{y,z}\right)  $ & \textbf{Physical} & $^{\left(  3\right)
}\mathbf{R}$ & \textbf{Spacetime} & \textbf{Stability}\\\hline
$B_{1}$ & $\left(  0,\Sigma,\Sigma\left(  \frac{1}{3}+\Sigma\right)  \right)
$ & $\left(  \sqrt{1-9\Sigma^{2}},0\right)  $ & Yes & $=0$ & $%
\begin{array}
[c]{c}%
\text{Bianchi (Kasner-like universe) for }\Sigma\neq0\\
\text{FLRW (Spatially flat) for }\Sigma=0
\end{array}
$ & Unstable\\
&  &  &  &  &  & \\
$B_{2}$ & $\left(  0,\Sigma,\Sigma\left(  \frac{1}{3}+\Sigma\right)  \right)
$ & $\left(  -\sqrt{1-9\Sigma^{2}},0\right)  $ & Yes & $=0$ & $%
\begin{array}
[c]{c}%
\text{Bianchi (Kasner-like universe) for }\Sigma\neq0\\
\text{FLRW (Spatially flat) for }\Sigma=0
\end{array}
$ & Unstable\\\hline\hline
\end{tabular}
\label{family2}%
\end{table}%

From the values of the parameters at the critical points we can extract
important information about the nature of the spacetime. As we discussed in
the previous sections there are two possible solutions which belong to the
Kantowski-Sachs and FLRW spacetimes. Hence, when the parameter $\Sigma$
vanishes, that is $\Sigma=0$, that is, the solution at the point has
$\sigma=0$, the resulting spacetime is FLRW, where the value of the spatial
curvature is calculated by the algebraic equation (\ref{ss.22}). Furthermore,
the Kantowski-Sachs solutions with $\Sigma\neq0$ are actually Bianchi I
spacetimes (Kasner-like universes) when $\Omega_{R}=0$.

In Table \ref{family1} the critical points of the first family of points (A)
are given, while the stability of the points is given. Similarly, Tables
\ref{family2} and \ref{family3} contain the points in families (B) and (C)
respectively. A discussion of the three families of critical points follows:

\begin{itemize}
\item \textit{Family A}: These critical points correspond to those of the
(original) Szekeres system (without the scalar field) and they were derived
earlier in \cite{silent1}. From the six physically accepted points, the
solutions at the points $A_{1}$ and $A_{2}$ correspond to those of FLRW
universe: point $A_{1}$ describes a dust solution, while $A_{2}$ describes the
Milne universe. The solution of the field equations at the points $A_{3}$ and
$A_{4}$ is described by the Kasner solutions of Bianchi type I spacetimes.
Furthermore, Kantowski-Sachs geometries correspond to the solutions at points
$A_{5}$ and $A_{6}$. From the study of the eigenvalues of the linearized
system close to the critical points we can extract information about the
stability of the points. We find that all the points of family (A) are unstable.

\item Family B: The critical points of this family are surfaces because the
parameter $\Sigma$ takes values in the interval $\frac{1}{3}\leq\Sigma
\leq\frac{1}{3}.$ The matter source at points $B_{1}$ and $B_{2}$ is that of a
stiff fluid and corresponds to the kinetic term of the scalar field when
$V\left(  \phi\right)  =0$. The parameter, $\varepsilon$, at these points
depends upon $\Sigma,$ as given by the expression $\varepsilon=\Sigma\left(
\frac{1}{3}+\Sigma\right)  .$ Moreover, we calculate that $\Omega_{R}=0$,
which means that for $\Sigma\neq0$ so the resulting solution is described by
the Bianchi universe, and actually for $\Sigma\neq0,-\frac{1}{3}$ the solution
is Kasner-like \cite{kasner111}, while for $\Sigma=-\frac{1}{3}$ the solution
reduces to the Kasner universe. Finally, when $\Sigma=0$, the resulting
solution is described by the spatially-flat FLRW universe with a stiff fluid.
These two points are always unstable.

\item Family C: The third family of critical points admits five physically
acceptable solutions, where the potential term and the kinetic part of the
scalar field contribute to the solution. Point $C_{1}$ describes a
spatially-flat FLRW universe with $\Omega_{D}\left(  C_{1}\right)  =0$ and
exists for values of $\lambda$ such that $\lambda^{2}\leq6,$ while the point
is stable when $\lambda^{2}<2$. \ At Point $C_{2}$ we have $\Omega_{D}\left(
C_{2}\right)  =1-\frac{3}{\lambda^{2}}$, which means that the solution exists
when $\lambda^{2}>3.$ Moreover the solution is described by a spatially-flat
FLRW universe while the solution is always unstable. The solution at point
$C_{3}$ is described by a FLRW geometry with non-zero spatial curvature, that
is $^{\left(  3\right)  }R=\frac{1}{3}\left(  \frac{4}{\lambda^{2}}-2\right)
$. Next, we find that when $\lambda^{2}>2$ the solution at the point $C_{3}$
is always stable, while $\Omega_{D}\left(  C_{3}\right)  =0$. Finally, the
solutions at the points $C_{4}$ and $C_{5}$ can describe a spatially-flat FLRW
geometry when $\lambda^{2}=2,$ a Kantowski-Sachs spacetime when $\lambda
^{2}>2,$ and a Bianchi III geometry when $\lambda^{2}<2.$ The stability
analysis shows that the solutions at these two critical points are always
unstable. From the latter points we can infer that the existence of a scalar
field gives solutions for the Szekeres system with positive spatial curvature.
\end{itemize}

%

\begin{table}[tbp] \centering
\caption{Critical points of family (C)}%
\begin{tabular}
[c]{ccccccc}\hline\hline
\textbf{Point} & $\left(  \mathbf{\Omega,\Sigma,\varepsilon}\right)  $ &
$\left(  \mathbf{y,z}\right)  $ & \textbf{Physical} & $^{\left(  3\right)
}\mathbf{R}$ & \textbf{Spacetime} & \textbf{Stability}\\\hline
$C_{1}$ & $\left(  0,0,0\right)  $ & $\left(  \frac{\lambda}{\sqrt{6}}%
,1-\frac{\lambda^{2}}{6}\right)  $ & Yes & $=0$ & FLRW (Spatially flat) &
Stable $\lambda^{2}<2$\\
$C_{2}$ & $\left(  1-\frac{3}{\lambda^{2}},0,0\right)  $ & $\left(
\sqrt{\frac{3}{2}}\frac{1}{\lambda},\frac{3}{2}\lambda^{2}\right)  $ & Yes &
$=0$ & FLRW (Spatially flat) & Unstable\\
$C_{3}$ & $\left(  0,0,0\right)  $ & $\left(  \sqrt{\frac{2}{3}}\frac
{1}{\lambda},\frac{4}{3\lambda^{2}}\right)  $ & Yes & $\neq0$ & FLRW (Spatial
curvature $\frac{1}{3}\left(  \frac{4}{\lambda^{2}}-2\right)  $) & Stable
$\lambda^{2}>2$\\
&  &  &  &  &  & \\
$C_{4}$ & $\left(  0,\frac{1}{6}-\frac{1}{2\left(  1+\lambda^{2}\right)
},\frac{\lambda^{2}-1}{6\left(  1+\lambda^{2}\right)  ^{2}}\right)  $ &
$\left(  \frac{\sqrt{\frac{3}{2}\lambda}}{1+\lambda^{2}},\frac{3}{2}%
\frac{2+\lambda^{2}}{1+\lambda^{2}}\right)  $ & Yes & $%
\begin{array}
[c]{c}%
\neq0~,~\lambda^{2}\neq2\\
=0~,~\lambda^{2}=2
\end{array}
$ & $%
\begin{array}
[c]{c}%
\text{Kantowski-Sachs for }\lambda^{2}>2\\
\text{Bianchi III for }\lambda^{2}<2\\
\text{FLRW (Spatially flat) for }\lambda^{2}=2
\end{array}
$ & Unstable\\
&  &  &  &  &  & \\
$C_{5}$ & $\left(  0,\frac{2-\lambda^{2}}{3+12\lambda^{2}},\frac{\lambda
^{2}\left(  \lambda^{2}-2\right)  }{2\left(  1+4\lambda^{2}\right)  ^{2}%
}\right)  $ & $\left(  \frac{3\sqrt{\frac{3}{2}\lambda}}{1+4\lambda^{2}}%
,\frac{9}{2}\frac{2+5\lambda^{2}}{1+4\lambda^{2}}\right)  $ & Yes & $%
\begin{array}
[c]{c}%
\neq0~,~\lambda^{2}\neq2\\
=0~,~\lambda^{2}=2
\end{array}
$ & $%
\begin{array}
[c]{c}%
\text{Kantowski-Sachs for }\lambda^{2}>2\\
\text{Bianchi III for }\lambda^{2}<2\\
\text{FLRW (Spatially flat) for }\lambda^{2}=2
\end{array}
$ & Unstable\\
$C_{6}$ & $\left(  -3-\frac{3}{\lambda^{2}},-\frac{1}{3},\frac{1}{6}\right)  $
& $\left(  \sqrt{\frac{3}{2}}\frac{1}{\lambda},\frac{3}{2}\lambda^{2}\right)
$ & No &  &  & \\
$C_{7}$ & $\left(  -\frac{3}{\lambda^{2}},\frac{1}{6},0\right)  $ & $\left(
\sqrt{\frac{3}{2}}\frac{1}{\lambda},\frac{3}{2}\lambda^{2}\right)  $ & No &  &
& \\\hline\hline
\end{tabular}
\label{family3}%
\end{table}%

At this point we want to discuss how the present analysis changes when we
consider a potential $V\left(  \phi\right)  $ different from the exponential
function. Technically in that consideration parameter $\lambda$ is not always
a constant hence we have a six-dimensional system to study. Therefore new
critical points can exists while the stability of critical points changes.
More specifically, for every $\lambda=\lambda_{0}$, such that $\Gamma\left(
\lambda_{0}\right)  =1$, then the rhs of equation (\ref{ss.21}) vanishes and
from the remain five equations we find the same critical points with that for
the exponential potential on the surface where $\lambda=\lambda_{0}$. However,
the new critical points which exists are those one where $y\rightarrow0$, in
order the rhs of (\ref{ss.21}) to be zero, and $z\neq0$. 

We find two possible points, with coordinates
\begin{align*}
\mathbf{D}_{1} &  :\left(  \Omega_{D_{1}},\Sigma_{D_{1}},\varepsilon_{D_{1}%
},y_{D_{1}},z_{D_{1}},\lambda_{D_{1}}\right)  =\left(  0,-\frac{1}{3},\frac
{1}{3},0,3,0\right)  ~~\text{and }^{\left(  3\right)  }\mathbf{R>0,}\\
\mathbf{D}_{2} &  :\left(  \Omega_{D_{2}},\Sigma_{D_{2}},\varepsilon_{D_{2}%
},y_{D_{2}},z_{D_{2}},\lambda_{D_{2}}\right)  =\left(  0,0,0,0,1,0\right)
~~\text{and }^{\left(  3\right)  }\mathbf{R=0.}%
\end{align*}
At these two points parameter $\lambda$ vanishes which means that $V_{,\phi
}\left(  \phi\right)  =0$. Therefore, the scalar field act as an cosmological
constant. The curvature at point $D_{1}$ has a positive value, hence
corresponds to the vacuum Bianchi III universe with cosmological constant
\cite{vacb3}. On the other hand, at point $D_{2}$ holds,\ $\Sigma_{D_{2}%
}=\varepsilon_{D_{2}}=0$, while $^{\left(  3\right)  }R=0$ which means that it
describes that this point describes the de Sitter universe. 

We do not continue with the stability analysis of the critical points because
it depends on the functional form of $\Gamma\left(  \lambda\right)  $,
consequently of the scalar field potential $V\left(  \phi\right)  $. 

\section{Conclusions}

\label{section5}

In this work we considered the Szekeres system for which we have assumed also
the existence of a scalar field. The scalar field is assumed to be homogeneous
so that the FLRW limit exists and the solutions of our dynamical system are
comparable with the Szekeres-Szafron spacetimes. We found that there are two
families of solutions which correspond to two different underlying spatial
geometries. More specifically, the line element which describes the geometry
can be that of the homogeneous Kantowski-Sachs spacetime or that of an
inhomogeneous FLRW(-like) spacetime. The later FLRW-like spacetimes are found
to be inhomogeneous, but their spatial curvature is not an arbitrary function
of one of the spatial variables as in the Szekeres system. Furthermore we find
that the constants of integration for the reduced system of the gravitational
field equations are constants and not functions of one of the space variables
as in the Szekeres geometries.

The second main difference with the Szekeres spacetimes is that the first
family of Kantowski-Sachs spacetimes are spatially homogeneous and not
inhomogeneous, that is, the first family of solutions are locally rotational
symmetric spacetimes which means that they admit a four-dimensional Killing
algebra, while the Szekeres spacetimes containing only dust do not admit any
Killing fields.

In order to study the stability of the two different spacetimes we performed
an analysis of the critical points for the gravitational field equations. In
order to perform that analysis we wrote the field equations in terms of the
kinematic quantities for the comoving observer, $u^{\mu}=\delta_{0}^{\mu}$,
and we normalized the parameters according to the $\theta$-normalization. In
order to perform our analysis we assumed the scalar field potential to be
exponential. We found three different families of critical points which
correspond to the (A) Szekeres system, (B) Szekeres system with stiff fluid
and (C) solutions where both the kinetic and potential parts of the scalar
field contribute.

The critical points of families (A) and (B) are found to be always unstable,
while there are only two possible stable solutions which belong to the third
family. Indeed, the possible stable solutions are points $C_{1}$ and $C_{3}$.
Point $C_{1}$ can describe an accelerated universe when it is stable, while
the solution at point $C_{3}$ is stable when it describes an open universe. On
the other hand, we found that there are two critical points, namely $C_{4}$
and $C_{5}$ which can describe solutions with a homogeneous Bianchi III geometry.

We conclude that the existence of an \textquotedblleft
inflaton\textquotedblright\ in the Szekeres system can lead to inhomogeneous
accelerated FLRW-like universes. Such an analysis is important in the
pre-inflationary epoch and our solutions extend the inhomogeneous de Sitter
generalizations of \cite{barcc}. In a forthcoming work we will generalize this
analysis to the Szekeres-Szafron system.

\begin{acknowledgments}
JDB is supported by the Science and Technology Facilities Council (STFC) of
the United Kingdom. AP acknowledges financial supported of FONDECYT grant No. 3160121.
\end{acknowledgments}

\end{document}